\documentclass[twocolumn,amsmath,amssymb,showpacs]{revtex4}
\usepackage{graphicx}
\usepackage{xcolor}

\begin{document}

\title{Impurity-induced resonant states in topological nodal-line semimetals}

\author{Tao Zhou$^{1,2}$}
\email{tzhou@scnu.edu.cn}
\author{Wei Chen$^{3}$}

\author{Yi Gao$^{4}$}

\author{Z. D. Wang$^{2}$}
\email{zwang@hku.hk}

\affiliation{$^{1}$Guangdong Provincial Key Laboratory of Quantum Engineering and Quantum Materials, GPETR Center for Quantum Precision Measurement, and School of Physics and Telecommunication Engineering,
South China Normal University, Guangzhou 510006, China\\
$^{2}$Department of Physics and Center of Theoretical and Computational Physics, The University of Hong Kong, Pokfulam Road, Hong Kong, China.\\
$^{3}$College of Science, Nanjing University of Aeronautics and Astronautics, Nanjing 210016, China.\\
$^{4}$Department of Physics and Institute of Theoretical Physics,
Nanjing Normal University, Nanjing, 210023, China.}

\date{\today}
\begin{abstract}
Nodal-line semimetals are characterized by a kind of topologically nontrivial bulk-band crossing, giving rise to almost flat
surface states. Yet, a direct evidence of the surface states is still lacking. Here we study theoretically impurity effects in topological nodal-line semimetals based on the T-matrix method. It is found that for a bulk impurity, some in-gap states may be induced near the impurity site, while the visible resonant impurity state can only exist for certain strength of the impurity potentials. For a surface impurity, robust resonant impurity states exist in a wide range of impurity potentials. Such robust resonant states stem from the topological protected weak dispersive surface states, which can be probed by scanning tunneling microscopy, providing a strong signature of the topological surface states in the nodal-line semimetals.

\end{abstract}
\pacs{74.90.+n, 71.90.+q, 03.65.Vf}
\maketitle

\section{introduction}
Three-dimensional topological nodal-line semimetals (NLSs) have recently attracted great interest in condensed matter physics and materials science~\cite{burk,ryu,kim,dwzhang,wchen1,jhu1,jhu,kum,hpan,wchen,wchen2,rong,
huh,volo,hir,lau,tak1,topp,neu,sch,mof,mof1,bian,xxu,qyu,quan,ywu,sato,tak,xie,chan,bsong,danwei,Bzdusek16,Chen17,Yan17,Ezawa17}. These NLSs are generally characterized by bands crossing closed loops near the Fermi level. Each loop carries a $\pi$ Berry flux, leading to the nontrivial surface states. The bulk and surface states may bring a number of interesting and important properties as well as potential applications, such as unique symmetric behavior~\cite{dwzhang}, surface chern insulator~\cite{wchen1}, nontrivial transport properties~\cite{jhu1,jhu,kum,hpan,wchen,wchen2}, giant Friedel oscillations~\cite{rong}, long-range Coulomb interaction~\cite{huh}, and possible route to high-Tc superconductivity~\cite{volo}.

Recently, the NLS has been proposed theoretically or reported experimentally  in various systems, including the Cu$_3$XN family (X=Ni, Cu, Pd, Ag, Cd)~\cite{ryu,kim}, alkali earth metals~\cite{rong,hir},  group-IV tellurides (SnTe, GeTe)~\cite{lau}, WHX family (W=Gd, Zr, Hf, La; H=Si, Ge, SN, Sb; X=O, S, Se, Te)~\cite{tak1,topp,neu,sch,jhu1,jhu,kum,hpan,mof,mof1}, XTaSe$_2$ family (X = Pb, Tl)~\cite{bian,xxu}, CaTX family (T = Cd, Ag; X = P, Ge, As), YX$_3$ family (Y=Ca, Sr, Ba;X=P, AS)~\cite{qyu,quan}, PtSn$_4$~\cite{ywu}, Ta$_3$SiTe$_6$~\cite{sato}, AlB$_2$~\cite{tak}, Ca$_3$P$_2$~\cite{xie,chan}, and the cold atom system~\cite{bsong,danwei}.
The research progress on the NLSs has made it possible to study further physical properties of these systems. On the other hand,
so far more evidences to hallmark the NLSs especially the surface states therein are still highly awaited.

The single impurity effect is an important tool for understanding  quantum states in various solid-state systems~\cite{bra}, which is rather powerful especially for the gapped systems.
The in-gap bound states may be induced by an impurity, through which some hidden properties may be detected.
 Theoretically, the impurity induced bound states are manifested in the local density of states (LDOS) near the impurity site~\cite{bra}.
 Experimentally, they can be measured through the scanning tunneling microscopy (STM)~\cite{oys}.
 Previously, the single impurity effects have been widely used to investigate physical properties such as the pairing symmetry in various unconventional
superconductors~\cite{bra,tsai,zhou,ola}. They have also been studied intensively in various topologically non-trivial systems, including several topological superconductor/superfluid systems~\cite{jsau,xjliu,nagai,wimm,hhu,nag,naga,zhou1,cha}, topological insulators ~\cite{rud,jlu},
and topological Weyl semimetals~\cite{hua}. The NLS is a relatively new member to topological materials.
Previously the quasiparticle interference in the normal and superconducting NLS was studied theoretically~\cite{cha}. And the
possible resonant state of the NLS with the $d$-wave superconducting order was also discussed~\cite{cha}.  
While to the best of our knowledge, the theoretical study on the single impurity effect in the normal state of NLS
is still lacking, which may provide an effective way to detect the unique properties of the NLS.

In this paper, we study theoretically the single impurity effect with different impurity potentials starting from an effective model to describe the NLS system. Given that the bulk and surface states in the NLS possess completely different properties, it can be expected that the impurity effect in the bulk and surface states should also lead to different results.
The LDOS near the impurity site is studied and
both the bulk and surface impurities are addressed. For a bulk impurity, some in-gap states may be induced near the impurity site, while the resonant impurity state can only exist for typical impurity potentials. In contrast, a resonant bound state induced by the surface impurity is revealed for a rather wide range of impurity potentials considered here.
We compare the results with that in the Weyl semimetal and show that the exotic surface states of the NLS can be detected
through the surface impurity effect.

The rest of the paper is organized as follows.
In Sec. II, we introduce
the model and present the relevant formalism. In Sec. III, we
report numerical calculations and discuss the obtained
results. Finally, we give a brief summary in Sec. IV.

\section{Model and formalism}

The NLS may contain one or more nodal lines with various configurations. For the latter case with nodal lines well separated in the reciprocal space, we assume that the impurity induced scattering between different nodal lines is negligibly small. Therefore, it is sufficient to consider the NLS carrying one nodal loop, which can be captured by the minimal model as
\begin{equation}\label{H}
H=(h-t k^2)\sigma_3+\lambda k_y \sigma_2.
\end{equation}
Here $\sigma_{2,3}$ are Pauli matrices in the two band space.
The eigenvalues of the Hamiltonian are
\begin{equation}
E_{\pm}=\pm\sqrt{(h-t k^2)^2+(\lambda k_y)^2},
 \end{equation}
which define a line node $k_x^2+k_z^2=h/t$ (in the $k_y=0$ plane).

In order to investigate the impurity effect, we introduce a tight-binding model of the Hamiltonian \eqref{H} on a cubic lattice as
\begin{eqnarray}\label{3D}
H=\sum_{\bf k}\varepsilon({\bf k})\sigma_0+\sum_{\bf k}M({\bf k})\sigma_3+\sum_{\bf k}\lambda({\bf k})\sigma_2,
\end{eqnarray}
where $\sigma_0$ is the $2\times2$ identity matrix, $\varepsilon({\bf k})=-2\sum\limits_{\alpha=x,y,z}t_{0\alpha}\cos k_\alpha-\mu$ is a diagonal term in addition to Eq. \eqref{H}, $M({\bf k})=-2\sum\limits_{\alpha=x,y,z}t_{1\alpha} \cos k_\alpha-h$, and $\lambda({\bf k})=2\lambda_0 \sin k_y$.
Then we obtain two energy bands from the above Hamiltonian as $E_\pm ({\bf k})=\varepsilon({\bf k})\pm \sqrt{M({\bf k})^2+\lambda({\bf k})^2}$. The first term breaks the chiral symmetry ($[H,\sigma_1]_+=0$) of the system and thus introduces finite dispersion to the surface states~\cite{burk}. Nevertheless, it will not change the band topology of the bulk states.

A point impurity term is given by
\begin{equation}\label{imp}
H_{imp}=V_s \sum_{\sigma} c^\dagger_{{\bf r_0}\sigma}c_{{\bf r_0}\sigma},
\end{equation}
where $V_s$ is the impurity potential. ${\bf r_0}$ is the site where the impurity locates.

To study the effect of the surface impurity, let us consider a sample with finite thickness in the $y$-direction with $1 \leq y \leq N_y$ ($N_y$ is the number of lattice sites in the $y$ direction) while periodic boundary condition still holds in the $x$ and $z$ directions.
Then the Hamiltonian is reduced to quasi-two-dimensional one in the $x-z$ plane by a partial Fourier transformation along the $y$-direction as
\begin{eqnarray}\label{2D}
H=& \sum\limits_{{\bf k},y,\sigma}\zeta_{{\bf k}\sigma}c^\dagger_{y\sigma}({\bf k})c_{y\sigma}({\bf k})\nonumber \\ & -(\sigma t_{1y}+t_{0y})\sum\limits_{{\bf k},y,\sigma}[c^\dagger_{y\sigma}({\bf k})c_{y+1,\sigma}({\bf k})+h.c.]\nonumber \\&+\sum\limits_{{\bf k},y}[\frac{\lambda}{2} c^{\dagger}_{{ y}\uparrow}({\bf k})c_{y+1\downarrow}({\bf k})-\frac{\lambda}{2} c^{\dagger}_{{y}\downarrow}({\bf k})c_{{y}+1\uparrow}({\bf k})+h.c.],
\end{eqnarray}
with the corresponding momentum being reduced to two-dimensional one as ${\bf k}={\bf k_{\parallel}}=(k_x,k_z)$ and $\zeta_{{\bf k}\sigma}=\sigma h-\mu-2\sum\limits_{\alpha=x,z}(\sigma t_{1\alpha}+t_{0\alpha})\cos k_\alpha$.

The Hamiltonian in Eqs. \eqref{3D} and \eqref{2D} can be written as the $2\times 2$ and $2 N_y\times 2 N_y$ matrix [$H=\sum_{\bf k}\Psi^{\dagger}({\bf k})\hat{M}\Psi({\bf k})$], with $\Psi({\bf k})=(c_{{\bf k}\uparrow},c_{{\bf k}\downarrow})^{T}$ and $\Psi({\bf k})=(c_{1\uparrow}({\bf k}),c_{1\downarrow}({\bf k}),\cdots, c_{N_y\uparrow}({\bf k}),c_{N_y\downarrow}({\bf k}) )^{T}$, respectively. In the following, we investigate the effects of both the bulk and surface impurities, in which periodic and open boundary conditions are adopted respectively.

The bare Green's function matrix
 $\hat{G}({\bf k},\omega)$ can be obtained
 by diagonalizng the Hamiltonian, with the matrix elements being expressed as,
\begin{equation}
G_{0ij}({\bf k},\omega)=\sum_n\frac{u_{i,n}({\bf k})u^{*}_{j,n}({\bf k})}{\omega-E^n({\bf k})+i\Gamma},
\end{equation}
with $u_{in}$ and $E^n$ being the eigen-vectors and eigen-values of the Hamiltonian matrix $H$, respectively.

The T-matrix and the full Green's function is expressed as,
\begin{eqnarray}
\hat{T}(\omega)&=&[\hat {I}-\hat{U}\hat {G_0}(\omega)]^{-1}\hat{U};\\
\hat{G}({\bf r},\omega)&=&\hat{G}_0(\omega)+\hat{G}_0({\bf r},\omega)\hat{T}(\omega)\hat{G}_0(-{\bf r},\omega).
\end{eqnarray}
Here, $\hat{G}_0(\omega)=\frac{1}{N}\sum_{\bf k}\hat {G_0}({\bf k},\omega)$ is the bare Green's function without the impurity,
$\hat{G}_0({\bf r},\omega)=\frac{1}{N}\sum_{\bf k}\hat {G_0}({\bf k},\omega)e^{i{\bf k}\cdot {\bf r}}$ is the Fourier transformation of $\hat{G}_0({\bf k},\omega)$ and $\hat{I}$ is the identity matrix. For a bulk impurity, $\hat{U}$ is a $2\times2$ matrix with $\hat{U}=V_s \hat{I}$ and for a surface impurity at the $y=1$ plane, $U$ is a $2N_y\times 2N_y$ matrix with two non-vanishing matrix elements $U_{11}=U_{22}=V_s$.
The LDOS can be calculated through
\begin{equation}
\rho({\bf r},\omega)=-\frac{1}{\pi}\mathrm{Im} [G_{11}({\bf r},\omega)+G_{22}({\bf r},\omega)].
\end{equation}

In the present work, we set the parameters to $t_{1\alpha}=1$, $\lambda=0.5$, and $h=-4$. The band degeneracy points define a nodal line
\begin{equation}
\cos k_x+\cos k_z=1;\quad  k_y=0.
\end{equation}
In the presence of the pseudo-spin independent hopping term with $\varepsilon({\bf k})\neq 0$, the surface states
are dispersive~\cite{burk}. In the following calculation, we set $t_{0x}=t_{0z}=0.2$, $t_{0y}=0.1$, and $\mu=-0.6$. Here, the chemical potential makes sure that the nodal line appears at the zero energy.

\section{Results and discussion}

  \begin{figure}
\centering
  \includegraphics[width=3.5in]{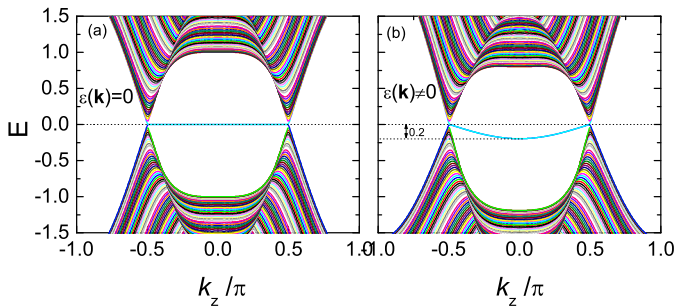}
\caption{(Color online) The energy spectrum along the $k_x=0$ cut with considering the open boundary condition along the $y$ direction and periodic boundary condition along the $x$ and $z$ directions.
}
\end{figure}
We first study the topological band structure considering the periodic boundary condition along the $x-z$ plane and the open boundary condition along the $y$ direction with $1\leq y\leq 200$.
The eigenvalues through diagonalizing Hamiltonian of Eq. \eqref{2D} as a function of $k_z$ with $k_x=0$ are plotted in Fig.~1. As $\varepsilon({\bf k})= 0$,
the Hamiltonian \eqref{3D} reduces to Eq. \eqref{H} in the continuous limit.
In this case, the energy spectrum is symmetric about zero energy due to chiral symmetry and a flat surface band appears at the Fermi energy, as seen in Fig.~1(a).
The existence of the flat surface band can be well understood through treating the in-plane momentum ${\bf k_\parallel}$ as effective parameters and the system reduces to be one-dimensional. One can define a topological invariant depending on ${\bf k_\parallel}$ to illustrate the topological properties~\cite{dwzhang},
\begin{equation}
C({\bf k_{\parallel}})=-\frac{i}{\pi}\int^{\pi}_{-\pi} \langle u_n({\bf k}) \mid \partial k_y \mid u_n({\bf k})\rangle d k_y.
\end{equation}
where $u_n({\bf k})$ is the eigen-vector of the valence band.
Then, we have $C=1$ when the in-plane momentum ${\bf k_{\parallel}}$ is inside the projection of the nodal loop,
and $C=0$ when ${\bf k_{\parallel}}$ is elsewhere [cf. Fig.~1(a)].
Therefore, when ${\bf k_{\parallel}}$ is inside the Fermi surface, there must be topological protected boundary states if the open boundary along the $y$-direction is considered. Without chiral-symmetry-breaking term, the energy of boundary modes do not vary with ${\bf k}_\parallel$ and lead to a flat band.
In real materials of NLS, there may exist a finite $\varepsilon({\bf k})$ term.
The energy spectrum in this case is plotted in Fig.~1(b). As is seen,
For this case, the chiral symmetry is broken and the surface band has a dispersion with the band width about 0.2
corresponding to the strength of $t_{0x}$ and $t_{0z}$.

We now discuss the impurity effect of NLSs. We first present the numerical results with $\varepsilon({\bf k})=0$.
Two types of the point impurity will be considered here. One is the bulk impurity inside a three-dimensional sample with periodic boundary condition in all the three directions. A point impurity at the site ${\bf r}=(0,0,0)$ is considered. The other is an impurity at the site ${\bf r}_\parallel=(0,0)$ on the open surface ($y=1$) of the sample where the periodic boundary condition holds only the $x$ and $z$ directions. In the calculation, the Hamiltonian in Eqs. \eqref{3D} and \eqref{2D} is adopted for the two cases respectively.
The LDOS spectra near a bulk impurity site [at the site (0,1,0)] with different impurity potentials are displayed in Fig.~2(a).
In the absence of impurity ($V_s=0$), the LDOS spectrum
is V-shaped with two peaks at the energies $\pm\Delta_0$, as indicated in Fig.~2(a). The shape of the bulk LDOS is similar to that of a nodal superconducting material and
 then the energy $2\Delta_0$ can be defined as an bulk energy gap.
 When the point impurity is added, some additional in-gap states appear (indicated by the arrow), with the energies depending on the impurity potential $V_s$.
As the impurity potential is weak ($V_s=5$), the in-gap state appears at the negative energy. When the potential increases to $V_s=10$, a sharp resonant peak appears near the zero energy. As $V_s$ increases to $15$, the in-gap peak shifts to the positive energy and the peak intensity decreases rapidly. As $V_s$ increases further, only a broad hump feature exists at a saturant positive energy.
Our numerical results indicate that for a bulk impurity in the NLS system, the in-gap states exist for all of the impurity intensities we considered. However, the impurity resonant state, behaving as a sharp in-gap peak, is not robust and exists only for a very small parametric range of the impurity potential.

  \begin{figure}
\centering
  \includegraphics[width=3.5in]{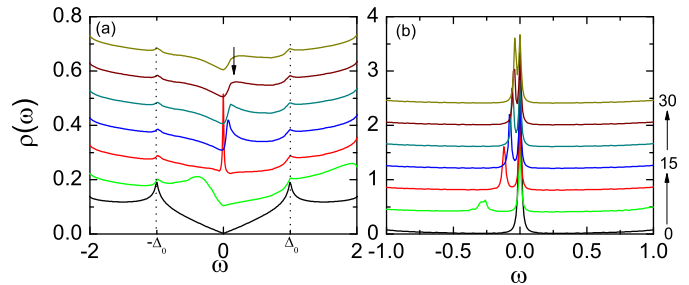}
\caption{(Color online) (a) The LDOS spectra near a bulk impurity with $\varepsilon({\bf k})=0$. (b) The LDOS spectra near a surface impurity with $\varepsilon({\bf k})=0$. From bottom to top, the impurity scattering potentials ($V_s$) are 0, 5, 10, 15, 20, 25, 30, respectively.
}
\end{figure}

We move to study the surface impurity effect. The LDOS spectra near a surface impurity [at the site $(0,1)$ in the $y=1$ plane] are plotted in Fig.~2(b). As the impurity potential is zero,
there is a sharp peak at the zero energy, due to the zero-energy flat band of the surface states. In the presence of the point impurity, an additional peak appears, with the energy moving towards zero energy as $V_s$ increases. One can see that the effect of a surface impurity is significantly
different from that of a bulk one. Here, the impurity-induced state is robust and exists below the Fermi energy for all values of the impurity potential. 

Now let us study the impurity effects with chiral-symmetry breaking, i.e., $\varepsilon({\bf k})\neq 0$.
The corresponding numerical results of the LDOS spectra are presented in Fig.~3.
Note that, the bare LDOS spectrum $(V_s=0)$ is asymmetric about zero energy ($\Delta_{-}\neq \Delta_{+}$), as is seen in Fig.~3, consistent with the band spectrum shown in Fig.~1(b). Therefore, both positive and negative impurity potentials will be considered in the following.

We first look into the case of positive impurity potentials. For the bulk LDOS spectra, the impurity effects remain qualitatively the same, i.e., in the presence of impurity term, there are some in-gap features, with the energy shifting from negative to positive when the impurity strengths become stronger. An
impurity-induced resonant peak appears near the zero energy for a typical potential $V_s=10$. For the surface spectra, In the absence of impurity, the low energy peak become wider, due to the dispersion of the surface band. The peak width (about 0.2) is consistent with the width of the surface band width shown in Fig.~1(b). In the presence of impurity, an additional resonant peak appears at the negative energy and approaches to the Fermi energy when the impurity strength increases. For rather strong potentials ($V_s\geq 20$), the impurity-induced resonant peak merges into the surface spectra and disappears. The impurity effects for $\varepsilon({\bf k})\neq 0$ are qualitatively similar to those for $\varepsilon({\bf k})= 0$, where the resonant peak induced by the surface impurity appears for a rather wide range of impurity potential and may be probed by the STM experiments.

We then present the numerical results of the negative impurity scattering potentials ($V_s<0$). For the bulk spectra, the in-gap states shift from the positive energy to negative energy as the potential ($\mid V_s \mid$) increases. A resonant peak appears for a typical potential $(V_s=-10)$. For the surface spectra, the impurity resonant peak appears at the positive energy and approaches to the Fermi energy for stronger potentials. For very strong potentials, the resonant peak disappears, the same as that for the positive potential. The difference of the positive impurity effect and negative impurity effect can be understood through a particle-hole transformation. The surface impurity induced resonant peak is robust for an impurity with moderate strong scattering potentials.

  \begin{figure}
\centering
  \includegraphics[width=3.5in]{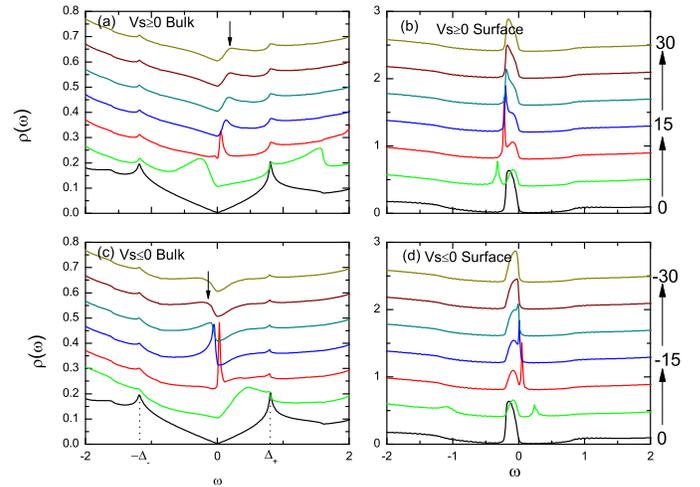}
\caption{(Color online) Similar to Fig.~2 while with $\varepsilon({\bf k})\neq0$ and both positive impurity potentials and negative ones are considered.
}
\end{figure}

The existence of the in-gap feature can usually be understood through analysing the denominator of the $T$-matrix term $[A(\omega)]$. Here the complex function $A(\omega)$ is expressed as
 \begin{equation}
A(\omega)=\mid \hat{I}-\hat{U}\hat{G}_0(\omega)\mid.
\end{equation}
The impurity resonance occurs when the pole condition $[A(\omega)\approx 0]$ satisfies.
Generally for the fully gapped system, the imaginary part $\text{Im} A(\omega)$ should tends to zero when the energy is smaller than the system gap. Then a resonance would occur if the real part $\text{Re} A(\omega)$ also equals to zero at certain energy levels. For the nodal system, $\text{Im} A(\omega)$ is usually finite and only tends to zero near the band crossing energy, where the resonance may occur.
Here, the bulk energy bands are not fully gapped and a nodal line exists. The resonance condition is difficult to achieve except for very low energies around the nodal line with vanishing DOS.
In contrast, for the case of surface impurity, the weight of the bare LDOS is dominated by the surface states and centers at zero energy. The LDOS spectra away from zero energy are suppressed greatly, so that the imaginary part $\text{Im} A(\omega)$ is rather small in a large energy scale, similar to gapped systems. As a result, a resonance may occur if the pole condition holds at this energy scale.
The above qualitative analysis partly explains our numerical results. 

  \begin{figure}
\centering
  \includegraphics[width=3.5in]{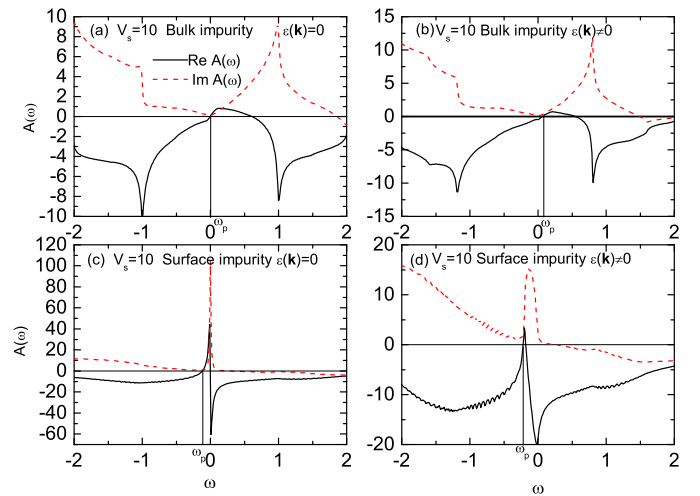}
\caption{(Color online) The real and imaginary parts of $A(\omega)$ as a function of the frequency $\omega$.
}
\end{figure}

We plotted the real and imaginary parts of the function $A(\omega)$ in Fig.~(4) to investigate and display in more detail the mechanism of the resonant states. For the bulk impurity with $V_s=10$, as seen in Figs.~4(a) and 4(b),
a pole occurs near the zero energy at the frequency $\omega_p$ when $\text{Re} A(\omega)$ equals to zero. At the same time, $\text{Im} A(\omega)$ approaches zero near the Fermi energy, and then a resonant behavior occurs.
For the surface impurity, as seen in Figs.~4(c) and 4(d), the pole condition can be satisfied in a wide range of the impurity potential, leading to the resonant impurity states. The surface resonant state is robust due to the weak dispersion of the surface states, which makes $\text{Im} A(\omega)$ exhibit a peak with finite width around zero energy. Correspondingly, the $\text{Re} A(\omega)$ shows logarithmic singularities, according to the
Kramers-Kronig relation~\cite{jxli}, as shown in Figs.~4(c) and 4(d). As a result, zero point of $\text{Re} A(\omega)$
generally exists around the band-crossing energy where $\text{Im} A(\omega)$ is very small. However, when the impurity potential becomes very strong, $\text{Re} A(\omega)$ is far below zero and the zero point does not exist. As a result, the pole condition cannot be satisfied.
In conclusion, the low energy resonant state generally exists for moderate strong surface impurity potential and disappears
for extremely strong impurity potential.

According to the above discussions, here the impurity-induced resonant states at the system surface are closely related to the weakly dispersive surface
states. The surface states are protected by the bulk band topology under the $\mathcal{PT}$-symmetry~\cite{dwzhang}. The impurity term [Eq. \eqref{imp}] obeys the $\mathcal{PT}$-symmetry and does not change the topological properties of the system. Therefore, the resonant states induced by the surface impurity is robust and exist for a wide range of impurity potentials. We expect that such a resonance peak can be observed in experiments which may provide a clear signature to identify the topological surface states in NLS materials.

\begin{figure}
\centering
  \includegraphics[width=3.5in]{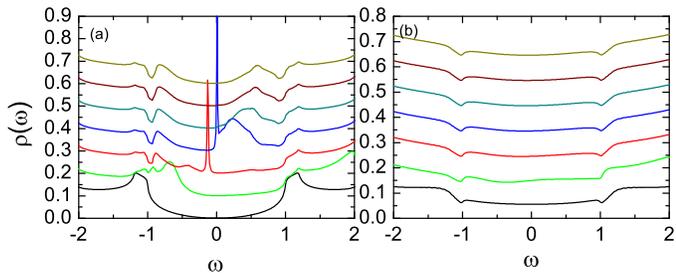}
\caption{(Color online) The LDOS spectra for the Weyl semimetal. (a) The LDOS spectra near a bulk impurity. (b) The LDOS spectra near a surface impurity. From bottom to top, the impurity potentials ($V_s$) are 0, 5, 10, 15, 20, 25, 30, respectively.
}
\end{figure}

It is worthwhile to investigate the impurity effect in the Weyl semimetal, and
compare it with that for topological NLS. A minimal lattice model to describe the topological Weyl semimetal can be
obtained by adding an additional term $k_x\sigma_1$ to Eq. \eqref{H}. The lattice model can be obtained
by adding the term $\sum_{\bf k} 2\lambda_0 \sin k_x \sigma_1$ to Eq. \eqref{3D} ($\varepsilon({\bf k})=0$ is adopted in the calculation).
Then, two point nodes $(0,0,\pm \pi/2)$ are obtained with the present parameters. At the system surface, a Fermi arc connecting the two nodal points can be obtained.

The impurity effects for the Weyl semimetal are studied in Fig.~(5).
The LDOS spectra at the nearest-neighbor of a bulk and the surface impurity are plotted in Figs.~5(a) and 5(b), respectively. As seen in Fig.~5(a), for the bulk impurity, the in-gap bound states seem to be qualitatively similar to those in NLSs, namely, sharp in-gap resonance peaks appear at near zero energy for certain impurity potentials. When the impurity strength is rather weak or strong, only an in-gap hump exists.
Note that here the resonant states can survive for a relatively wider range of scattering strength, compared with the bulk impurity effects of NLSs.
On the other hand, for the surface impurity, the results for Weyl semimetal are significantly different from those for NLSs. As  seen in Fig.~5(b), there are no in-gap features for all of the potentials we considered. Unlike the zero energy peak of the bare LDOS in the NLS, the Fermi arc surface states distribute in a finite energy scale and lead to a nearly constant LDOS. As a result, the resonant state cannot form. We also compare the results of three-dimensional topological insulators. It was reported that robust bound states can be induced by a bulk impurity~\cite{rud,jlu}, while the impurity bound states cannot form at the system surface~\cite{jlu}.
From the discussion above, one can see that the existence of the sharp impurity resonant states at the system surface can serve as a notable feature for topological NLSs and may be used to identify the NLS experimentally.

\section{summary}
In summary, single impurity effects in nodal-line semimetals have been symmetrically studied based on the T-matrix method. For a bulk impurity, there are some in-gap features for all of the impurity potentials we considered. While the impurity resonant states only exist near the Fermi energy for certain strength of impurity potential. For the surface impurity, there is a resonant impurity state for a rather wide range of impurity potentials. We also compare our results with those in other three-dimensional topologically non-trivial materials and conclude that the existence of the robust resonant impurity state at the system surface may provide a useful probe for the topological surface states in nodal-line semimetals.

\begin{acknowledgments}
	 This work was supported by the NKRDP of China (Grant No. 2016YFA0301800), and the GRF (Grants No. HKU 173309/16P and No. HKU173057/17P) and a CRF (No. C6005-17G) of Hong Kong.
\end{acknowledgments}


\end{document}